\documentclass[12pt]{article}
\usepackage{amssymb,amsmath,bm,bbm}
\usepackage{epsf}
\usepackage{epsfig}
\usepackage{afterpage}
\usepackage{longtable}
\usepackage{cite}

\def\simge{\mathrel{\rlap{\raise 0.511ex \hbox{$>$}}{\lower 0.511ex \hbox{$\sim$}}}}
\def\simle{\mathrel{\rlap{\raise 0.511ex \hbox{$<$}}{\lower 0.511ex \hbox{$\sim$}}}}
\def\slash#1{\setbox0=\hbox{$#1$}\dimen0=\wd0
      \setbox1=\hbox{/} \dimen1=\wd1 \ifdim\dimen0>\dimen1
      \rlap{\hbox to \dimen0{\hfil/\hfil}} #1                        \else
      \rlap{\hbox to \dimen1{\hfil$#1$\hfil}}
      /   \fi}

\newcommand{\lsim}{
\mathrel{\hbox{\rlap{\hbox{\lower4pt\hbox{$\sim$}}}\hbox{$<$}}}}

\newcommand{\gsim}{
\mathrel{\hbox{\rlap{\hbox{\lower4pt\hbox{$\sim$}}}\hbox{$>$}}}}

\newcommand{\vcb}{|V_{cb}|}

\newcommand{\tev}{\, {\rm TeV}}
\newcommand{\gev}{\, {\rm GeV}}
\newcommand{\mev}{\, {\rm MeV}}

\newcommand{\Heff}{{\cal H}_\text{eff}}

\newcommand{\mtb}{\overline{m}_{\rm t}}
\newcommand{\mcb}{\overline{m}_{\rm c}}

\newcommand{\mw}{M_{\rm W}}

\allowdisplaybreaks[2]

\newcommand{\be}{\begin{equation}}
\newcommand{\ee}{\end{equation}}
\newcommand{\bea}{\begin{eqnarray}}
\newcommand{\eea}{\end{eqnarray}}
\newcommand{\nn}{\nonumber}
\newcommand{\bi}{\begin{itemize}}
\newcommand{\ei}{\end{itemize}}

\def\kpn{K^+\rightarrow\pi^+\nu\bar\nu}

\def\klpn{K_{L}\rightarrow\pi^0\nu\bar\nu}

\begin{document}
\begin{titlepage}
\vspace*{-0.5truecm}

\begin{flushright}
TUM-HEP-666/07\\
MPP-2007-33
\end{flushright}

\vfill

\begin{center}
\boldmath

{\Large\textbf{Littlest Higgs Model with T-Parity\\
\vspace{0.2truecm}
 Confronting the New Data on $D^0-\bar D^0$ Mixing}}

\unboldmath
\end{center}

\vspace{0.4truecm}

\begin{center}
{\bf Monika Blanke$^{a,b}$, Andrzej J.~Buras$^a$,  Stefan Recksiegel$^a$,\\
Cecilia  Tarantino$^a$ and Selma Uhlig$^a$}
\vspace{0.4truecm}

 $^a${\sl Physik Department, Technische Universit\"at M\"unchen,
D-85748 Garching, Germany}

 {\sl $^b$Max-Planck-Institut f{\"u}r Physik (Werner-Heisenberg-Institut), \\
D-80805 M{\"u}nchen, Germany}
\end{center}
\vspace{0.6cm}
\begin{abstract}
\vspace{0.2cm}
\noindent 
Motivated by the first experimental evidence of meson oscillations in the $D$
system, we study $D^0 - \bar D^0$ mixing in the Littlest Higgs model
with T-parity, we investigate its role in constraining the model parameters
and its impact on the most interesting flavour observables.
We find that the experimental data are potentially strongly constraining but at
present limited by large theoretical uncertainties in the long-distance Standard Model
contribution to $D^0 - \bar D^0$ mixing.
\end{abstract}

\vfill\vfill
\end{titlepage}

\thispagestyle{empty}

\begin{center}
{\Large\bf Note added}
\end{center}

\noindent
An additional contribution to the $Z$ penguin in the Littlest Higgs model with T-parity has been pointed out in \cite{Goto:2008fj,delAguila:2008zu}, which has been overlooked in the present analysis. This contribution leads to the cancellation of the left-over quadratic divergence in the calculation of some rare decay amplitudes. Instead of presenting separate errata to the present work and our papers \cite{Blanke:2006eb,Blanke:2007db,Blanke:2007wr,Blanke:2008ac} partially affected by this omission, we have presented a corrected and updated analysis of flavour changing neutral current processes in the Littlest Higgs model with T-parity in \cite{Blanke:2009am}.

\newpage

\setcounter{page}{1}
\pagenumbering{arabic}
\noindent
The phenomenon of meson-antimeson oscillation is very sensitive to heavy
degrees of freedom propagating in the mixing amplitude and, therefore,
represents one of the most powerful probes of New Physics (NP).
In $K$ and $B_{d,s}$ systems the comparison of observed meson mixing with the
Standard Model (SM) prediction has achieved a good accuracy and plays a
fundamental role in constraining not only the unitarity triangle but also
possible extensions of the SM.
The evidence for flavour oscillation of the charmed meson
$D^0$, instead, has been reported only very recently by
BaBar \cite{babarDD} and Belle \cite{belleDD}, independently.
These experimental results have been combined in \cite{Ciuchini:2007cw} to which we refer for
details.
Here we just mention that the analysis in \cite{Ciuchini:2007cw} allows for CP-violation in
mixing but not in the decay amplitudes where the SM tree-level contributions
are expected to dominate.
CP-violation in $D^0 - \bar D^0$ mixing is also strongly suppressed, in the SM,
by the small combination of CKM matrix elements $V^{}_{cb} V_{ub}^*$.
In the presence of NP, however, new CP-violating contributions can occur and spoil
this feature.

Combining the new BaBar and Belle measurements of $D^0 - \bar D^0$ mixing
parameters yields, in particular,  an improvement
of almost an order of magnitude on the $\Delta M_D$ constraint, which now reads
\cite{Ciuchini:2007cw} 
\be
{\Delta M_D = (11.7 \pm 6.8)  \cdot 10^{-3}} \,\text{ps}^{-1}\,.
\label{eq:DMDexp}
\ee
This first evidence of $D^0 - \bar D^0$ mixing is certainly welcome as,
involving mesons with up-type quarks, it is complementary to mixing in $K$ and
$B_{d,s}$ systems in providing information on NP.
In order to discover a signal for NP in $\Delta M_D$, however, one would need
high confidence that the SM predictions lie well below the present
experimental limit.

Unfortunately, the SM calculation of $\Delta M_D$ is plagued by long-distance
contributions, responsible for very large theoretical uncertainties. 
In fact, unlike $B_{d,s}^0 - \bar B_{d,s}^0$ mixing that is completely dominated by 
short-distance effects generated by the top quark, in $\Delta M_D$ the
non-perturbative physics associated with long-distance effects
(e.\,g. propagation of light intermediate states) is potentially large and may
even dominate over the short-distance ones \cite{reviews}.

The short-distance contribution in $\Delta M_D$ \cite{Cheng:1982hq,Datta:1984jx}, indeed, is highly suppressed
both by a factor $(m_s^2-m_d^2)/M_W^2$ generated by the GIM mechanism and by a
further factor $(m_s^2-m_d^2)/m_c^2$ due to the fact that the external
momentum, of the order of $m_c$, is communicated to the internal light quarks
in box-diagrams. 
These factors explain why the box-diagrams are so small for $D$ mesons
relative to $K$ and $B_{d,s}$ mesons where the GIM mechanism enters as
$m_c^2/M_W^2$ and $m_t^2/M_W^2$ and external momenta can be neglected.
Moreover, a recent study of $\Delta M_D$  has found that NLO (QCD)
corrections to short-distance contributions interfere destructively 
with the LO ones, with the net effect $(\Delta M_D)_\text{SM}^\text{SD} \simeq
2 \cdot 10^{-6} \,\text{ps}^{-1}$ \cite{Golowich:2005pt}.

Within the SM, then, the short-distance contribution is negligible and a
reliable theoretical prediction requires an accurate knowledge of
the long-distance ones, whose estimates follow at present two approaches.
The ``inclusive'' approach, based on the
operator product expansion (OPE), relies on local quark-hadron duality and on
$\Lambda_\text{QCD}/m_c$ being small enough to allow a truncation of the
series after the first few terms.
The charm mass, however, may not be large enough for such an approximation.
The ``exclusive'' approach, on the other hand, sums over intermediate
hadronic states, which can be modeled or fit to experimental data.
These exclusive contributions, however, need to be known with high precision
due to cancellations between states within a given $SU(3)$ multiplet and,
furthermore, the $D^0$ is not light enough that its decays are dominated by few
final states.
As a consequence, in the absence of sufficiently precise data, some
assumptions are required and yield quite model-dependent results.

While most studies of long-distance contributions find $(\Delta
M_D)_\text{SM}^\text{LD} \lsim 10^{-3} \,\text{ps}^{-1}$, values as high as 
$10^{-2} \,\text{ps}^{-1}$ cannot be excluded {\cite{Bigi,Exclusive,Falk:2001hx}}.
The latter estimates, being of the order of magnitude of the experimental
constraint in \eqref{eq:DMDexp}, presently prevent revealing an unambiguous sign
of NP in $D^0 - \bar D^0$ mixing.
In spite of that, in view of 
future theoretical improvements as well as better experimental accuracies, it is certainly interesting to study possible NP
contributions to $\Delta M_D$ in specific extensions of the SM.
It is important to note that NP contributions appear in box-diagrams with
internal new heavy particles and, therefore, are of short-distance only.
In predictive NP models, a reliable calculation is then possible and
its remaining uncertainty is dominated by the parameters of the model itself.

The aim of the present Letter is to study the phenomenon of $D^0 - \bar D^0$ mixing in
the Littlest Higgs model with T-parity (LHT) and to investigate its impact on
our previous LHT flavour analyses \cite{BBPTUW,Blanke:2006eb}.

The LHT model \cite{ACKN,CL} belongs to the class of the so-called Little Higgs models (LH) \cite{ACG}, whose
basic idea for solving the {\it little hierarchy problem} is the interpretation
of the Higgs as a pseudo-Goldstone boson of a spontaneously broken global symmetry. 
Diagrammatically, the quadratic divergences that affect the
Higgs mass, within the SM, are canceled by the contributions of new heavy particles having
the same spin-statistics as the SM ones and masses around $1 \tev$.
In the LHT model, a discrete symmetry called T-parity is added, in order to 
satisfy the electroweak precision constraints \cite{HMNP}, 
by avoiding tree-level contributions of the new heavy gauge bosons and 
restoring the custodial $SU(2)$ symmetry.
Under T-parity particle fields are T-even or T-odd. 
The T-even sector consists of the SM particles and a heavy top
$T_+$, while the T-odd sector contains heavy gauge bosons ($W_H^\pm,Z_H,A_H$),
a scalar triplet ($\Phi$) and the so-called mirror fermions, i.\,e.
fermions corresponding to the SM ones but with opposite T-parity and 
$\mathcal{O}(1 \tev)$ mass.
Mirror fermions are characterized by new flavour violating interactions with SM fermions
and heavy gauge bosons, which involve two new unitary 
mixing
matrices in the quark sector, analogous to the Cabibbo-Kobayashi-Maskawa (CKM) matrix $V_\text{CKM}$.
They are $V_{Hd}$ and
$V_{Hu}$, when the SM quark is of down- or up-type respectively,
and they satisfy $V_{Hu}^\dagger V^{}_{Hd}=V^{}_\text{CKM}$ \cite{HLP}.
A similar structure is valid for the lepton sector as discussed in details in \cite{Blanke:2007db}.
It is important to recall two important features of the LHT model in order to
understand its role in flavour physics.
The first is that, because of these new mixing matrices, the LHT model does not belong to the Minimal
Flavour Violation (MFV) class of models whether constrained \cite{UUT} or
general \cite{AMGIISST} and significant 
effects in flavour observables are possible.
The second is that no new operators, and no new non-perturbative
uncertainties, in addition to the SM ones appear in the
LHT model.

Extensive flavour physics analyses in the LHT model have been recently
performed in both the quark {\cite{HLP,BBPTUW,Blanke:2006eb,SHORT,Hong-Sheng:2007ve,Blanke:2007wr}} and lepton sector \cite{Blanke:2007db,Indian}.
In particular, $D^0 - \bar D^0$ mixing has been studied in \cite{HLP,BBPTUW}, before it
was experimentally observed.
Motivated by the improved experimental constraint on $\Delta M_D$ \cite{babarDD,belleDD,Ciuchini:2007cw} we
update and extend here the LHT analysis of $D^0 - \bar D^0$ mixing.

As discussed above, at present the large SM long-distance uncertainties 
prevent to reveal an unambiguous NP contribution to $\Delta M_D$.
We choose, therefore, to disentangle our analysis from the large SM
uncertainties.
To this end, we consider only the LHT
contribution to $\Delta M_D$ and determine the range of values that it can
assume once the known flavour constraints are
imposed as in \cite{BBPTUW,Blanke:2006eb}.
Once the SM uncertainties are significantly reduced, our strategy can be
pushed further to use the experimental $\Delta M_D$ measurement to constrain
the parameters of the LHT model.
Moreover, if the smaller SM upper bounds $(\Delta M_D)_\text{SM}^\text{LD} \lsim 10^{-3} \,\text{ps}^{-1}$ are
confirmed, it will be legitimate to neglect the SM contributions and to use the
experimental $\Delta M_D$ measurement as a constraint for the LHT contribution
only.  

Meson-antimeson mixing in the LHT model is discussed in details 
in {\cite{BBPTUW,HLP}} where the separate contributions of T-even and T-odd sector to the
off-diagonal dispersive matrix elements $M_{12}^i$ ($i=K,d,s$ for $K$ and
$B_{d,s}$ systems) are explicitely given.
{
The $D$ system presents a difference since it involves external SM up-type
quarks and  therefore the T-even $T_+$ cannot run in the loop, so that the T-even
contribution reduces to the SM one. In the T-odd sector, both down-type mirror quarks, together with the charged gauge bosons $W_H^\pm$, and up-type mirror quarks, together with the neutral gauge bosons $Z_H, A_H$, contribute. 

Due to the near equality of up- and down-type mirror quark masses the formula for the $\Delta C=2$ effective Hamiltonian can straightforwardly be obtained from the one describing $\Delta S=2$ transitions calculated in \cite{HLP,BBPTUW}, yielding
\be
[\Heff(\Delta C=2)]_\text{odd} = 
\frac{G_F^2}{64\pi^2}M_{W}^2\frac{v^2 }{f^2}\eta_D
\sum_{i,j}\xi^{(D)}_i\xi^{(D)}_j F_H(z_i,z_j)\,(\bar uc)_{V-A}(\bar uc)_{V-A}\,.
\ee
Here, $z_i = {m_{Hi}^2}/{M_{W_H}^2}$ with $m_{Hi}$ denoting the mass
of the $i$-th mirror quark doublet, and the function $F_H$ has been
{determined} in \cite{HLP,BBPTUW}. The only difference with
respect to $[\Heff(\Delta S=2)]_\text{odd}$ is the fact that now the
relevant quark mixing is given by the $V_{Hu}$ matrix, leading to {the combination} $\xi^{(D)}_i$} \footnote{We would like to caution the reader that in \cite{BBPTUW} the indices of the $V_{Hu}$ elements have been erroneously interchanged.}
\be
\xi^{(D)}_i = V_{Hu}^{*iu} V_{Hu}^{ic}\,.
\ee
The QCD correction $\eta_D$ can be approximated by \cite{eta2B}
\be
\eta_D\simeq \eta_2 = 0.57\pm0.01\,.
\ee

The mirror quark contribution to the off-diagonal element $M_{12}^D$ of the neutral $D$-meson mass matrix is then found to be
\bea\nn
\left(M_{12}^D\right)_\text{odd}
&\equiv &\left|\left(M_{12}^D\right)_\text{odd}\right|\,e^{-2i\phi_D}\\
&=&
\frac{G_F^2}{48\pi^2}F_D^2\hat B_D m_{D^0} M_{W}^2 \frac{v^2}{f^2} \eta_2 \sum_{i,j}{\xi^{(D)}_i}^*{\xi^{(D)}_j}^* F_H(z_i,z_j)\,.
\eea
Our definition of the phase $\phi_D$ follows from
\be\label{eq:6}
(M_{12}^D)^* = \langle \bar D^0 | \Heff(\Delta C=2) | D^0 \rangle \equiv
\left|\left(M_{12}^D\right)\right|\,e^{2i\phi_D}\,,
\ee 
where we note that it is $(M_{12}^D)^*$ and not $M_{12}^D$, as sometimes found
in the literature, that {appears on the l.\,h.\,s.~of \eqref{eq:6}}.
We stress that the theoretical uncertainty on the LHT contribution, being of
short-distance origin only, comes from the non-perturbative uncertainties in the decay
constant $F_D$ and the $B$-parameter $\hat B_D$, in addition to the new LHT
parameters scanned over in the analysis.
For $F_D$ we use the recent experimental determination by the CLEO-c
collaboration \cite{CLEO-c} that turned out to be in agreement with recent
lattice calculations \cite{domainwall,Aubin:2005ar} and of comparable precision.
Concerning the parameter $\hat B_D$, we consider the result of the most recent
(quenched) lattice calculation \cite{domainwall}, compatible within quite large
uncertainties with an older lattice determination \cite{Gupta:1996yt}.
Their numerical values are collected together with the other inputs in Table
\ref{tab:input}, where some numbers have been updated with respect to \cite{BBPTUW,Blanke:2006eb}.
\begin{table}[!]
\renewcommand{\arraystretch}{1}\setlength{\arraycolsep}{1pt}
\center{\begin{tabular}{|l|l|}
\hline
{\small $G_F=1.16637\cdot 10^{-5} \gev^{-2}$} & {\small$\Delta M_K= 3.483(6)\cdot 10^{-15}\gev$} \\
{\small$\mw= 80.425(38)\gev$} & {\small$\Delta M_d=0.508(4)/ \rm{ps}$\hfill\cite{BBpage}} \\\cline{2-2}
{\small$\alpha=1/127.9$} &{\small $\Delta M_s = 17.77(12)/\text{ps}$\hfill\cite{CDFnew,D0}} \\\cline{2-2}
{\small$\sin^2 \theta_W=0.23120(15)$\qquad\hfill\cite{PDG}} & {\small
  $F_K\sqrt{\hat B_K}= 143(7)\mev$\qquad\hfill\cite{Hashimoto,PDG}}\\\hline
{\small$|V_{ub}|=0.00409(25)$} &  {\small $F_D\sqrt{\hat B_D}= 241(24)\mev$\hfill\cite{CLEO-c,domainwall}}\\\cline{2-2}
{\small $\vcb = 0.0416(7)$\hfill\cite{BBpage}} & {\small$F_{B_d} \sqrt{\hat B_{B_d}}= 214(38)\mev$} \\\cline{1-1}
{\small$\lambda=|V_{us}|=0.2258(14)$ \hfill\cite{CKM2005}} & {\small$F_{B_s} \sqrt{\hat B_{B_s}}= 262(35)\mev$\;\;\hfill\cite{Hashimoto}} \\\hline
 {\small$\gamma=82(20)^\circ$ \hfill\cite{UTFIT}} & {\small$\eta_1=1.32(32)$\hfill\cite{eta1}} \\\hline
{\small$m_{K^0}= 497.65(2)\mev$} & {\small$\eta_3=0.47(5)$\hfill\cite{eta3}}\\\cline{2-2}
{\small$m_{D^0}=  1.8645(4)\gev$} &{\small$\eta_2=0.57(1)$} \\
{\small$m_{B_d}= 5.2794(5)\gev$} & {\small$\eta_B=0.55(1)$\hfill\cite{eta2B}}\\\cline{2-2}
{\small$m_{B_s}= 5.370(2)\gev$} & {\small$\mcb= 1.30(5)\gev$} \\
{\small $|\varepsilon_K|=2.284(14)\cdot 10^{-3}$ \hfill\cite{PDG}} &{\small$\mtb= 161.7(20)\gev$} \\
\cline{1-1}
{\small $S_{\psi K_S}=0.675(26)$ \hfill\cite{BBpage}} & \\
\hline
\end{tabular}  }
\caption {\textit{Values of the experimental and theoretical
    quantities used as input parameters.}}
\label{tab:input}
\renewcommand{\arraystretch}{1.0}
\end{table}

Having at hand all the LHT formulae for meson oscillations in $K$, $D$ and $B_{d,s}$
systems and rare $K$ and $B$ decays presented in \cite{BBPTUW,Blanke:2006eb} and here, we will now investigate the
impact of the measurement \eqref{eq:DMDexp} and of the constraints on
$|M_{12}^D|$ and $\phi_D$ derived in \cite{Ciuchini:2007cw} on the
parameters of the LHT model and our results presented in
\cite{BBPTUW,Blanke:2006eb}.
To this end we will consider two frameworks for the SM contributions to 
$D^0 - \bar D^0$ mixing:

\vspace{0.2cm}
\underline{\bf Framework X}

\noindent
The SM contribution to $M_{12}^D$ is set to zero so that the constraint on
$|M_{12}^D|$ and the phase $\phi_D$ shown in the {lower left} plot in Fig.~2 of
\cite{Ciuchini:2007cw} is directly applied to the LHT contribution.

\underline{\bf Framework Y}

\noindent
The SM contribution is allowed to vary within its large uncertainties as done
in \cite{Ciuchini:2007cw} and the general constraint on the NP contribution
shown in the {lower right} plot in Fig.~2 of \cite{Ciuchini:2007cw} is applied to the 
LHT model. 

\begin{figure}[t]
\begin{minipage}{8.0cm}
\center{\epsfig{file=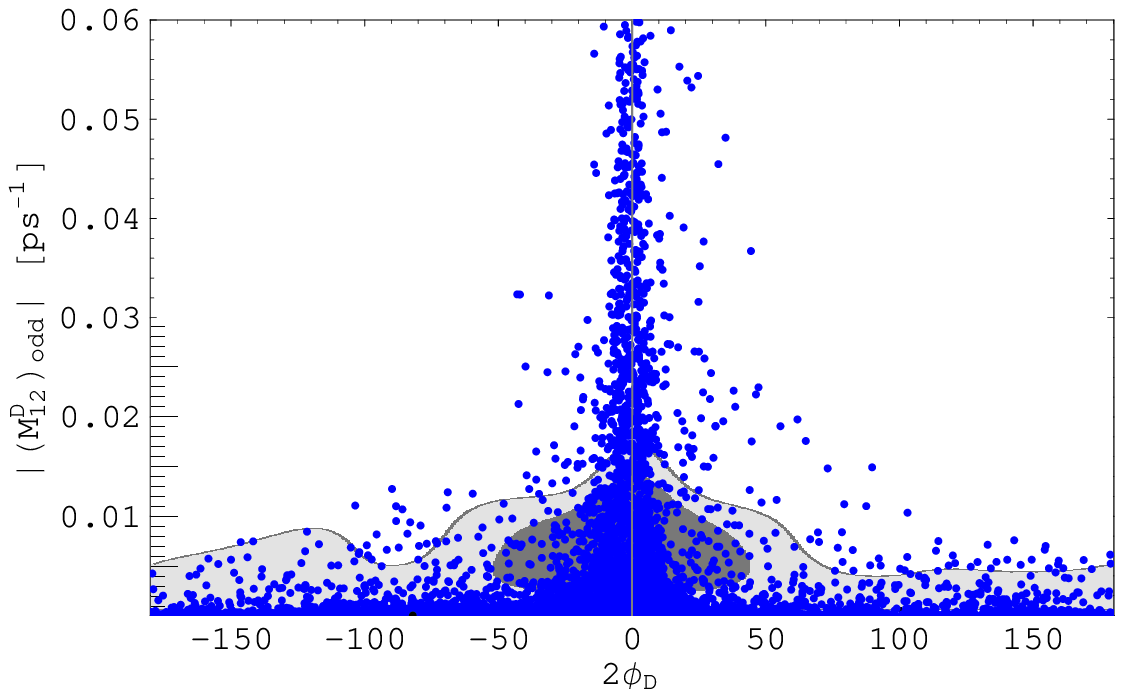,scale=0.65}}
\end{minipage}
\begin{minipage}{8.0cm}
\center{\epsfig{file=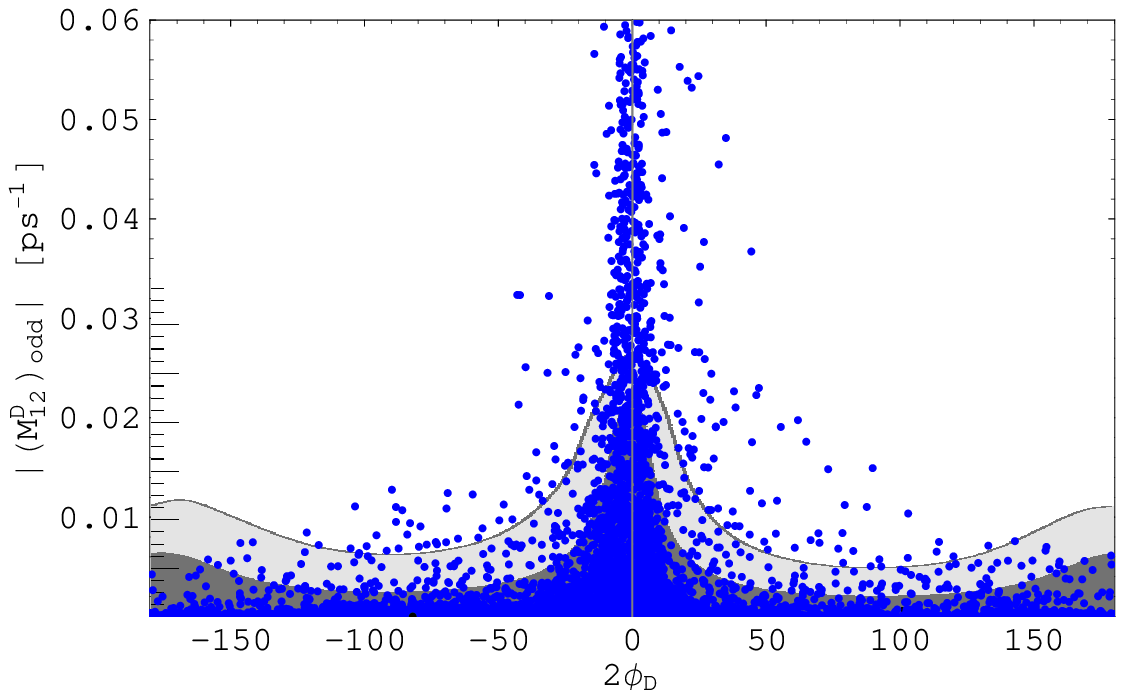,scale=0.65}}
\end{minipage}
\caption{\it $|M_{12}^D|$ versus $2\phi_D$ from a general scan over the 
LHT parameters, compared to the probability density function derived in
\cite{Ciuchini:2007cw}, for the Framework X (left) and Y (right).}
\label{fig:constrscan}
\end{figure}

\begin{figure}
\begin{minipage}{8.0cm}
\center{\epsfig{file=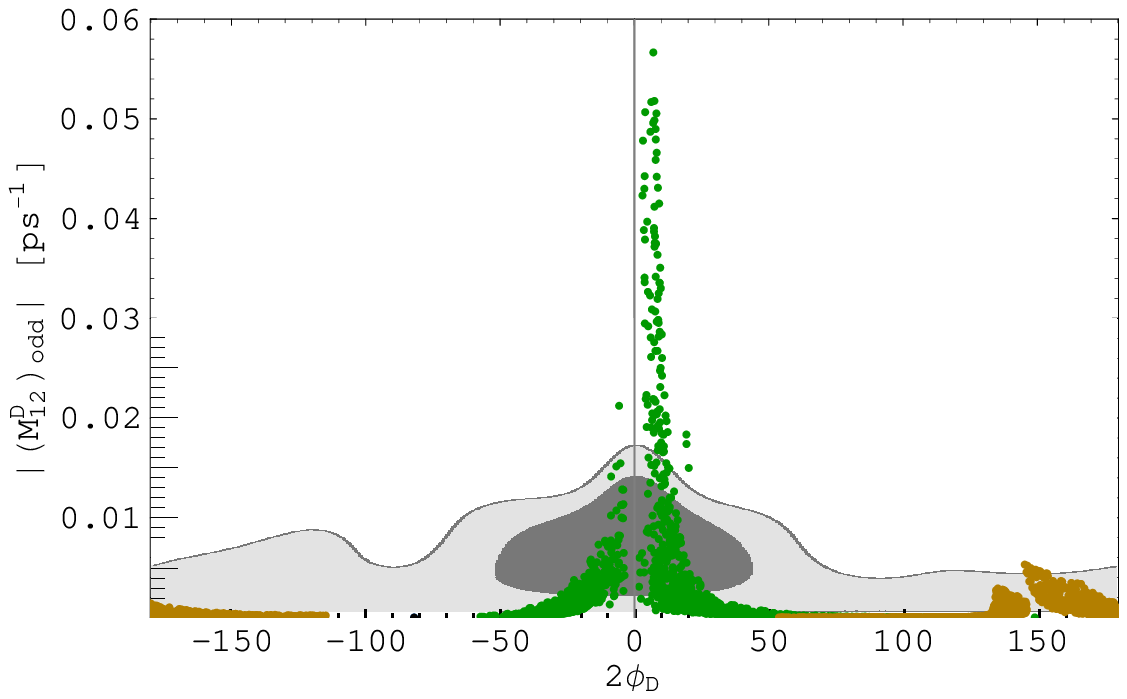,scale=0.65}}
\end{minipage}
\begin{minipage}{8.0cm}
\center{\epsfig{file=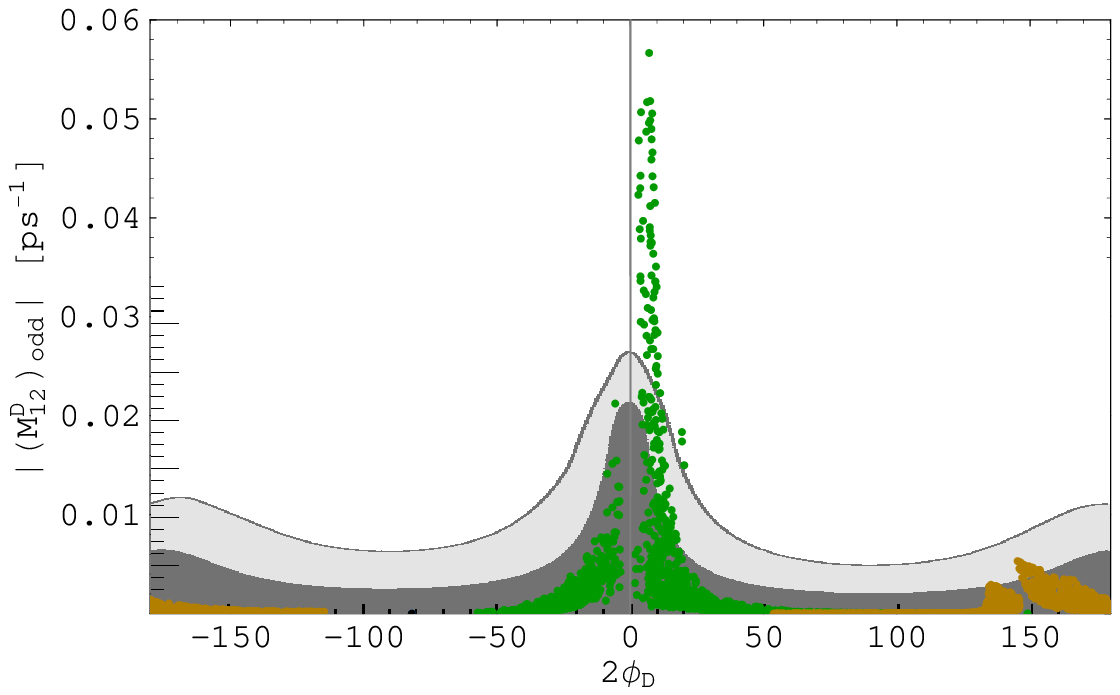,scale=0.65}}
\end{minipage}
\caption{\it The same as in Fig.~\ref{fig:constrscan} but for two LHT 
specific scenarios: $K$-Scenario (brown points) and $B_s$-Scenario 
(green points).}
\label{fig:constrKB}
\end{figure}

\begin{figure}
\begin{minipage}{8.0cm}
\center{\epsfig{file=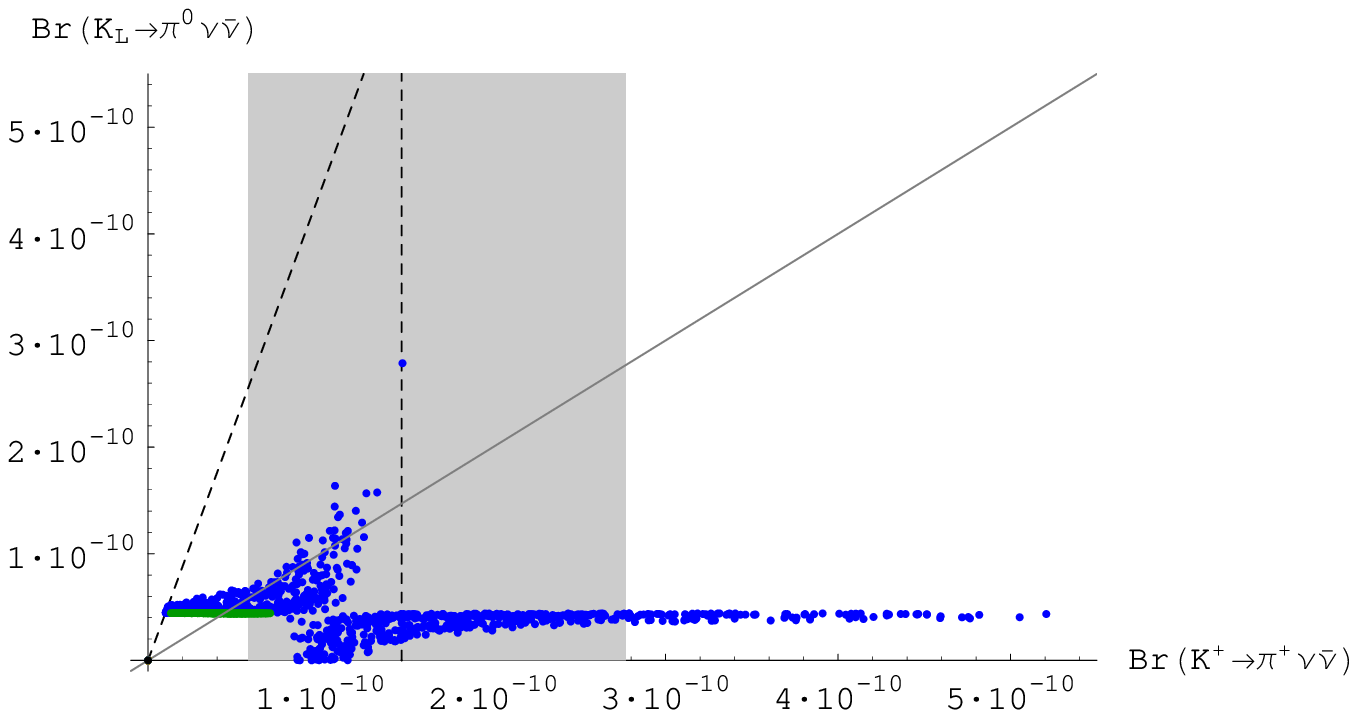,scale=0.55}}
\end{minipage}
\begin{minipage}{8.0cm}
\center{\epsfig{file=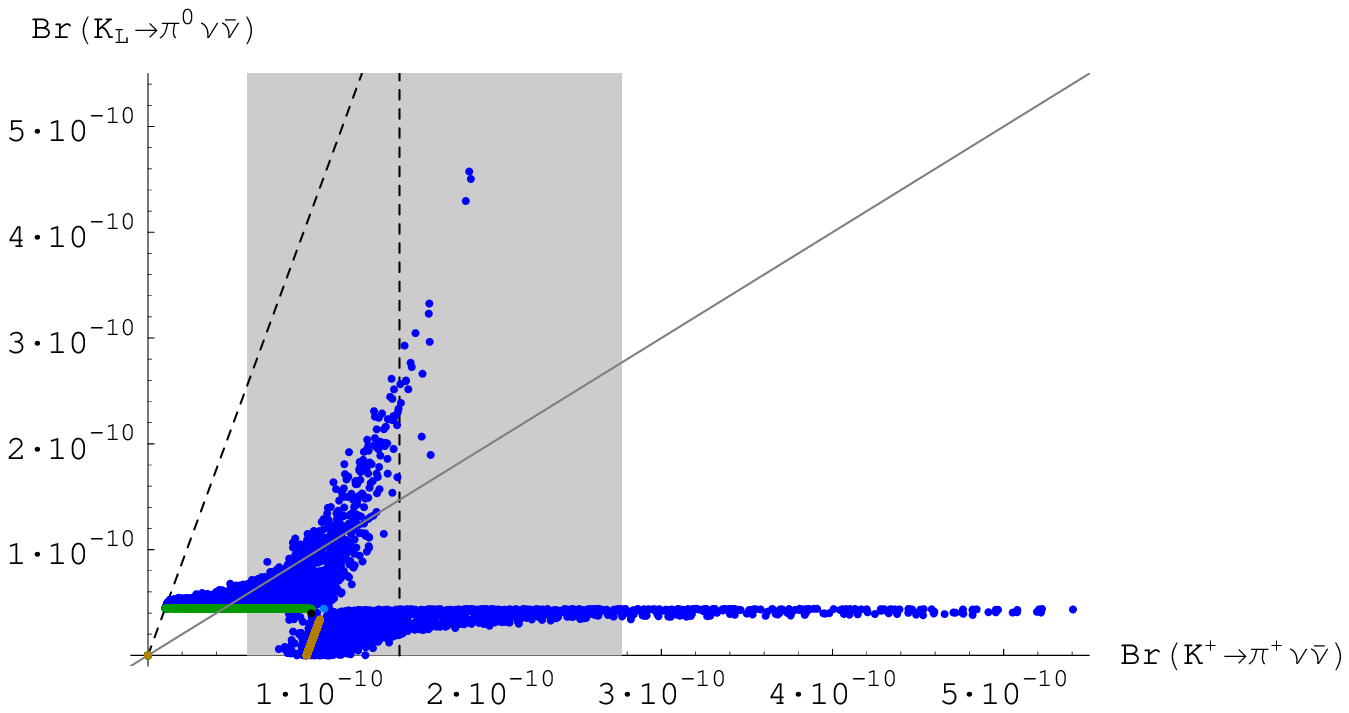,scale=0.55}}
\end{minipage}
\caption{\it $Br(\klpn)$ as a function of $Br(\kpn)$, after applying the
  {$1\sigma$-constraint} on $D^0 - \bar D^0$ mixing within the Framework X (left) and Y (right). The shaded
    area represents the experimental $1\sigma$-range for $Br(\kpn)$. The
    GN-bound \cite{GNbound} is displayed by the dotted line, while the solid line
    separates the two areas where $Br(\klpn)$ is larger or smaller than
    $Br(\kpn)$. }
\label{fig:KLKp}
\end{figure}

\begin{figure}
\begin{minipage}{7.5cm}
\center{\epsfig{file=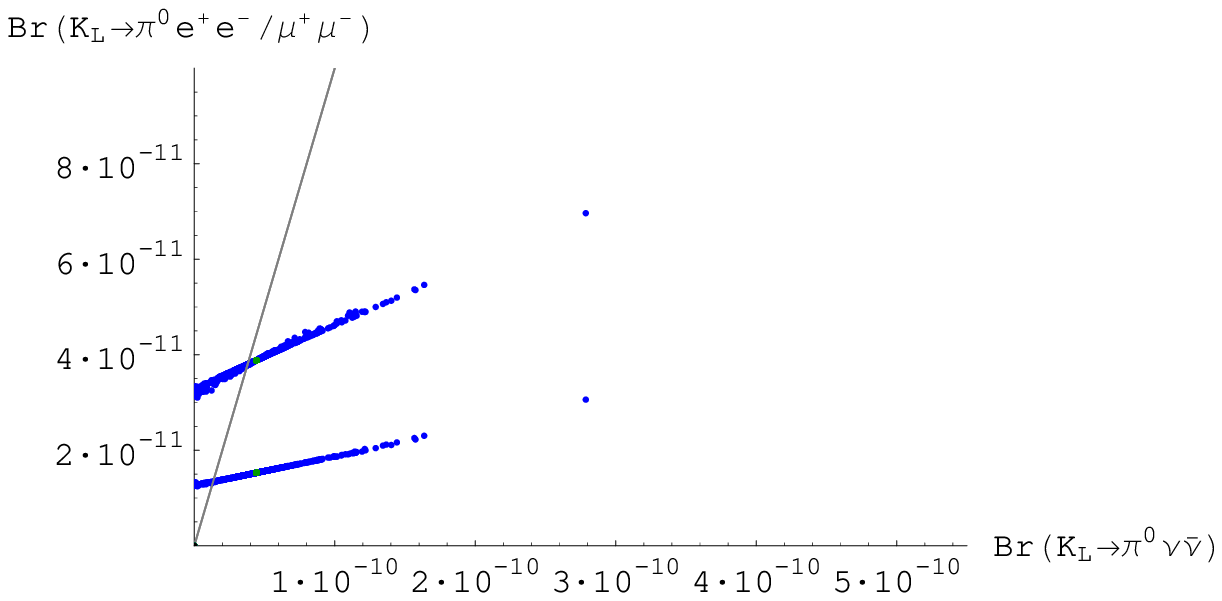,scale=0.62}}
\end{minipage}
\begin{minipage}{7.5cm}
\center{\epsfig{file=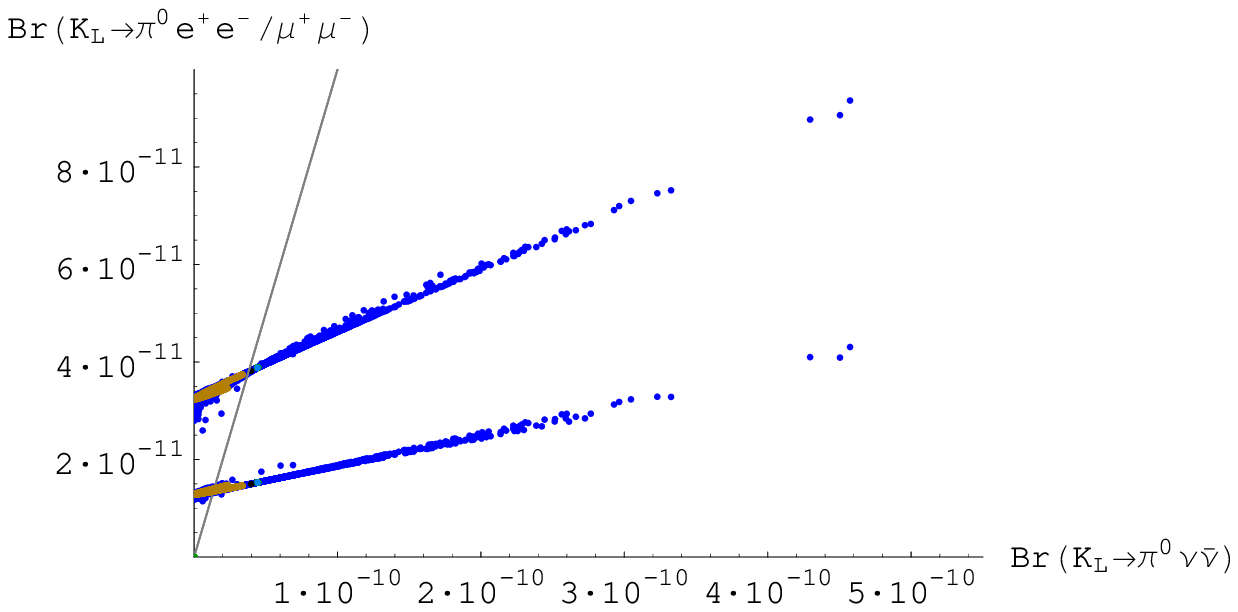,scale=0.62}}
\end{minipage}
\caption{\it $Br(K_L\to \pi^0 e^+e^-)$ (upper curve) and  $Br(K_L \to \pi^0
  \mu^+\mu^-)$ (lower curve) as functions of $Br(\klpn)$, after applying the
  {$1\sigma$-constraint} on $D^0 - \bar D^0$ mixing within the
  Framework X (left) and Y (right). }
\label{fig:KLKll}
\end{figure}

\begin{figure}
\begin{minipage}{8.0cm}
\center{\epsfig{file=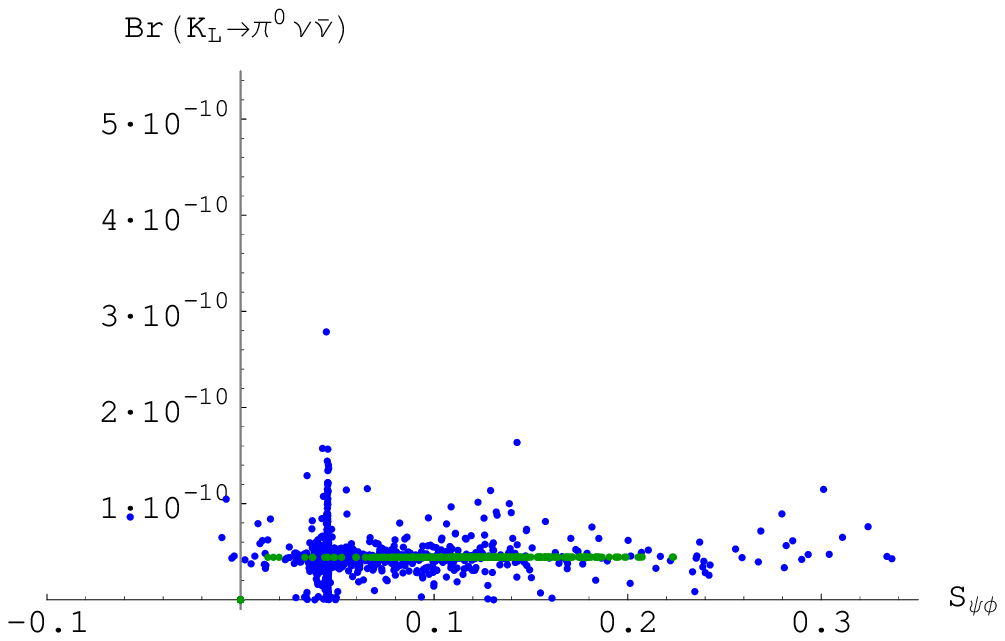,scale=0.65}}
\end{minipage}
\begin{minipage}{8.0cm}
\center{\epsfig{file=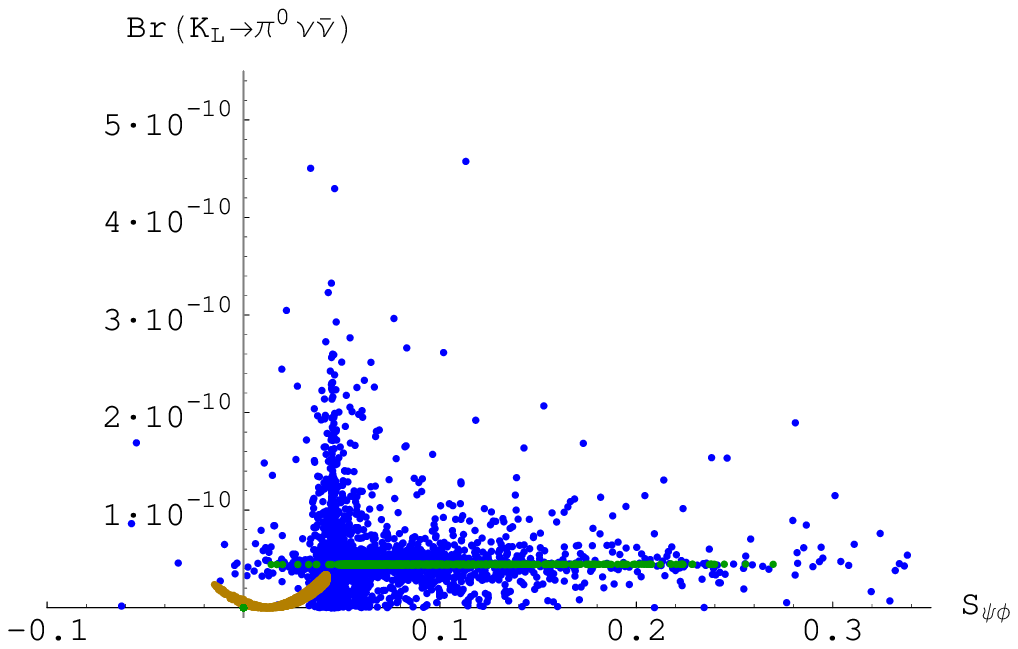,scale=0.65}}
\end{minipage}
\caption{\it $Br(K_L \to \pi^0
\nu \bar \nu)$ as a
  function of $S_{\psi \phi}$, after applying the
   {$1\sigma$-constraint} on $D^0 - \bar D^0$ mixing within the Framework X (left) and Y
  (right).}
\label{fig:KLSpsiphi}
\end{figure}

In Fig.~\ref{fig:constrscan} we show the predictions of the LHT model for
$|M_{12}^D|$ and $2 \phi_D$ obtained in a general scan (blue points) over the parameters
of the model in comparison with the allowed {$1\sigma$
  ranges}\footnote{In
  practice the constraints derived in \cite{Ciuchini:2007cw} have been
  implemented in our analysis, in the X and Y Frameworks respectively, 
  {approximating} the $1\sigma$ ranges as: X) $0.0025\,\text{ps}^{-1} \leq|M_{12}^D| \leq 0.0125\, \text{ps}^{-1}$
  and $2 |\phi_D| \leq 50^\circ$; Y) $|M_{12}^D| \leq 0.005 \,\text{ps}^{-1}$ or
  ($2 |\phi_D| \leq 25^\circ$ and $|M_{12}^D| \leq 0.02
 \, \text{ps}^{-1}$).} derived in
\cite{Ciuchini:2007cw}.
{If we allowed for the $2\sigma$ ranges instead, there would be almost no restrictions
on the Little Higgs parameter space from $D^0-\bar D^0$ mixing,
i.\,e. there {would be} no
visible difference between the plots with the points allowed in frameworks
X and Y. We therefore restrict ourselves to the $1\sigma$ ranges where the
effect is quite pronounced. {If in the future the $2\sigma$ ranges 
come down} to where there are now the $1\sigma$ ranges, $D^0-\bar D^0$
mixing and in particular its CP-violating phase will
put significant restrictions on the Little Higgs parameter space.}
Fig.~\ref{fig:constrKB} shows analogous results in two specific parameter
scenarios identified in \cite{BBPTUW,Blanke:2006eb}: the $K$-scenario (brown points) and
$B_s$-scenario (green points) that lead to large departures from the SM in $K$ and
$B$ decays, respectively.
Finally in Figs.~\ref{fig:KLKp}-\ref{fig:KLSpsiphi}, we show the impact of
the experimental $D^0 - \bar D^0$ constraint on the most interesting results
found in \cite{BBPTUW,Blanke:2006eb}.

From the inspection of Figs.~\ref{fig:constrscan}-\ref{fig:KLSpsiphi} and the
comparison with our previous results \cite{BBPTUW,Blanke:2006eb} we learn that:
\begin{itemize}
\item{The $D^0 - \bar D^0$ constraint is much weaker in the Framework Y, due to
    very large long-distance uncertainties present in the SM. In the Framework X
    the latter are only present in $F_D\sqrt{\hat B_D}$ and, as seen in
    Figs.~\ref{fig:constrscan}-\ref{fig:constrKB}, the impact of the $D^0 -
    \bar D^0$ constraint on the points satisfying all remaining observables is
    rather significant.}
\item{We observe from Fig.~\ref{fig:constrKB} that whereas the $K$-scenario is
  practically excluded by the $D^0 - \bar D^0$ mixing data in the
  Framework X, the impact on the $B_s$-scenario is only moderate in both SM frameworks.
Therefore, the impact of the $D^0 - \bar D^0$ constraint turns out to be
  significantly larger on $K$ decays than $B$ decays.
The reason is that both $K$ decays and $D^0 - \bar D^0$ mixing
  describe transitions between the first two quark generations, thus involving
  the same combinations of elements of $V_{Hd}$ and $V_{Hu}$, respectively.
Now, as $V_{Hd}$ and $V_{Hu}$ are related via $V^{}_{Hu}=V^{}_{Hd}
  V_\text{CKM}^\dagger$ and $
{V_\text{CKM}\simeq \mathbbm{1}}$, it approximately
  turns out that $V_{Hu} \simeq V_{Hd}$. Therefore, the observed correlation
  between $K$ and $D$ physics in indeed expected within the LHT model.}
\item{As shown in Figs.~\ref{fig:KLKp}-\ref{fig:KLSpsiphi}, in the case of
    the Framework Y and a general scan over the LHT parameters, very large
    departures from the SM expectations for rare $K$ decays and $S_{\psi \phi}$
    are possible. These plots, in fact, are qualitatively similar to those
    presented in \cite{BBPTUW,Blanke:2006eb}.}
\item{On the other hand, if mirror fermion contributions describe the full $D^0
    - \bar D^0$ mixing as supposed in the Framework X, the enhancements of rare $K$ decay
    branching ratios are significantly smaller than those found in
    \cite{Blanke:2006eb}, although they can still be substantial.
For instance $Br(K_L \to \pi^0 \nu \bar \nu)$ can be larger by a factor $5$
relative to the SM prediction. On the other hand the CP-conserving decay
  $Br(K^+ \to \pi^+ \nu \bar \nu)$ is less affected by the $D^0 - \bar D^0$
  mixing constraint.}
\item{
Finally we observe that a large phase $\phi_D$ can be generated, signaling the possibility of sizeable CP-violating effects in the $D$ meson system within the LHT model. Quantitative predictions for CP-violating observables in the $D$ system would however require a much more detailed analysis which is beyond the scope of the present Letter. }
\end{itemize}

The main message of our paper is that the present data on $D^0 - \bar D^0$
mixing put already significant constraints on the predictions of the LHT model
for $K$ and $B$ decays.
However, without a consistent improvement in the estimate of the SM
contribution to $D^0 - \bar D^0$ mixing, the role of the $D$ system in
constraining the parameters of the LHT model as well as other extensions of
the SM will be limited, even if the accuracy of the data improves. {The situation is more promising in the case of CP-violation in
the $D$ meson system, where due to the absence of SM contributions
much cleaner predictions in a given NP model can be made. On the
other hand, useful constraints on the parameter space of a given NP
model can only be obtained once the data significantly improve.}

\subsection*{Acknowledgements}
{We thank Ikaros Bigi for an illuminating and informative discussion on the physics of $D^0-\bar D^0$ oscillations.} This research was partially supported by {the Cluster of Excellence `Origin
and  Structure of the Universe' and by} the German Bundesministerium f{\"u}r 
Bildung und Forschung under contract 05HT6WOA.

\end{document}